\documentclass[a4paper,fleqn]{cas-dc}
\usepackage{hyperref}
\usepackage[capitalise,noabbrev]{cleveref}
\usepackage{bm}
\usepackage{caption}
\usepackage{float}
\usepackage{placeins}
\usepackage{cuted}
\usepackage{multirow}
\usepackage{braket}
\usepackage[numbers,square]{natbib}
\usepackage{algorithm}
\makeatletter
\newcommand{\figurehere}{\def\@captype{figure}}
\newcommand{\tablehere}{\def\@captype{table}}
\makeatother
\setcitestyle{square,sort&compress}
\AtBeginDocument{\let\cite\citep}
\usepackage{tabularx}
\newcolumntype{Y}{{\raggedright\arraybackslash}X}

\begin{document}
\let\WriteBookmarks\relax

\newcolumntype{L}[1]{>{\raggedright\arraybackslash}p{#1}}
\newcolumntype{R}[1]{>{\raggedleft\arraybackslash}p{#1}}
\newcolumntype{C}[1]{>{\centering\arraybackslash}p{#1}}
\renewcommand{\topfraction}{.95}
\renewcommand{\bottomfraction}{.95}
\renewcommand{\textfraction}{.05}
\renewcommand{\floatpagefraction}{.75}
\setcounter{topnumber}{5}
\setcounter{bottomnumber}{5}
\setcounter{totalnumber}{10}

\shorttitle{Apple Vision Pro: The good, the bad, the ugly}


\title [mode = title]{Apple Vision Pro: \textit{The good, the bad, the ugly}}
\author[1]{Santiago Berrezueta-Guzman}[orcid=0000-0001-5559-2056]
\credit{Conceptualization, Writing, Supervision}

\author[1]{Andrei Koshelev}[orcid=0009-0000-3786-4994]

\author[1]{Habiba-Loai Khalil}[orcid=0009-0007-3085-5105]

\author[1]{Stefan Wagner}[orcid=0000-0002-5256-8429]

\affiliation[1]{organization={Technical University of Munich},city={Heilbronn}, country={Germany}}



\begin{abstract}
Apple Vision Pro (AVP), the headset Apple introduced as its first ``spatial computer,'' built by a single, vertically integrated vendor, entered the market in February 2024 and has since accumulated a fast-growing body of primary research spanning surgery, education, accessibility, design, and human-computer interaction.
In this paper, we adopt the tripartite structure of Leone's 1966 film to organize a systematic review of this literature: we cataloged the validated, working applications of AVP (the good), the persistent technical and ergonomic limitations of a still-early-generation headset (the bad), and the ethical concerns and structural harms that follow from deliberate design, marketing, and deployment choices (the ugly). 
The review is built from a Scopus export of 218 records mentioning AVP, which we narrowed to 104 papers (2024--2026) with a primary focus on AVP, using explicit inclusion/exclusion criteria that we report in full, alongside a PRISMA-informed flow diagram of the screening process.
We find the corpus heavily skewed toward validating benefits (medicine and surgery, education, and software tooling account for over half of the included papers) and comparatively thin on documented risks or misuse, a pattern we treat as a finding in its own right rather than a limitation to be explained away. 
We further show that several recurring issues, namely price, simulated social presence, and the gaze-tracking privacy architecture, resist a clean separation into ``bad'' or ``ugly'' using the published evidence alone, since both readings remain compatible with the same data. 
We close by identifying where the evidence base most needs to grow: independent risk and ethics research, bystander- and equity-focused study designs, and disclosure that would let researchers distinguish an engineering backlog from a structural choice.
\end{abstract}


\maketitle

\section{Introduction}\label{I}

When Sergio Leone's \textit{The Good, the Bad, and the Ugly} \cite{114} premiered in 1966, its tripartite structure captured a simple idea: that the same unfolding situation can contain genuine value, unresolved difficulty, and moral compromise at once, without those three registers collapsing into one another. 
We borrow that structure in this study for a concrete case in spatial computing: \textit{Apple Vision Pro (AVP)}, the mixed-reality (MR) headset Apple introduced as its first ``spatial computer'' \cite{120}, a term predating the product by two decades.

AVP is a useful case for this kind of review because, unlike ``artificial intelligence (AI),'' a label that spans many companies, architectures, and constantly shifting systems, AVP names one company's product, running one operating system on one hardware generation at a time. 
Its enumerable set of components includes a forward-facing camera and depth-sensor array, an eye-tracking pipeline, an outward-facing ``EyeSight'' display and ``Persona'' avatar system, a curated App Store, and a launch price of \$3,499 \cite{121}. 

Apple updated the hardware once during the period this review covers, releasing an M5-chip model on 22 October 2025 with a redesigned, counterbalanced ``Dual Knit Band,'' a GPU that renders 10\% more pixels on the same micro-OLED displays, and passthrough refresh rates up to 120Hz, at an unchanged \$3,499 starting price; Apple's own materials describe this as an upgraded Vision Pro rather than a numbered second generation, though third-party coverage commonly calls it that \cite{112}. 

Since its release, a fast-growing body of academic research has evaluated what this specific device does well, where it still falls short, and what it enables people to do to one another, deliberately or otherwise; \Cref{sec:framework} quantifies the growth and composition of this literature within our own included corpus.

This paper makes two contributions. 
First, we conduct a systematic screening of the AVP literature indexed in Scopus, documented in Section~\ref{sec:method} with a PRISMA-informed flow diagram, and organize what survives screening into a good/bad/ugly catalog: Section~\ref{sec:good} covers the good (validated, working applications), Section~\ref{sec:bad} the bad (technical and ergonomic limitations still being iterated on), and Section~\ref{sec:ugly} the ugly (ethical concerns and structural harms arising from deliberate choices). Each section closes with a summary table.
 
Second, we identify genuine overlaps between the bad and the ugly catalogs. Price, simulated social presence, and the gaze-tracking privacy architecture each admit two readings that the published evidence cannot adjudicate between.

This paper also presents, in \Cref{sec:discussion}, what this indeterminacy, together with the shape of the current evidence base, implies for where AVP research should go next, and, in \Cref{sec:conclusion}, concludes this study.

\section{Review Method}\label{sec:method}

The reviewed literature was identified and screened, and the included papers were subsequently classified into the good, bad, and ugly catalog that follows.

\subsection{Literature search strategy}\label{sec:search}

We searched Scopus for records mentioning \texttt{Apple Vision Pro} in \textit{Title}, \textit{Abstract}, or \textit{Keywords}, and exported all matches, yielding 214 records. We exported the results twice. The first export, on 10 September 2026, initially returned 242 records; a corrected re-export returned 214. We did not preserve a record of why the two exports differed, but none of the 28 dropped records is part of the included corpus. 
An updated export on 9 October 2026 contained the same 214 records plus four newly indexed ones, for a total of 218. 
All screening reported below was performed on this updated export.

\subsection{Literature inclusion/exclusion criteria}\label{sec:incexc}

We screened every record's title, abstract, and document type against four criteria, applied by close reading rather than keyword matching:

\begin{enumerate}
\item Document type: we \textit{excluded} Letters, Notes, Conference reviews, Books, and Book chapters, retaining Articles, Conference papers, Reviews, and Data papers for further screening; 
\item Primary focus: a record was retained only if AVP itself, rather than headsets in general, was the central subject of its stated contribution (a paper benchmarking AVP alongside Meta Quest 3 and HoloLens 2 as one of several comparison devices was excluded on this ground even when methodologically sound); comparative studies were retained where AVP was the device under evaluation and other devices served only as baselines against which AVP's performance was measured, and excluded where AVP appeared as one of several devices being surveyed with no single device central to the paper's contribution; 
\item Language and genre: we excluded records whose abstract text was not in English and pure opinion or editorial pieces with no study, system, or evaluation behind them; and \item Completeness: records with no usable abstract were excluded (those that only mention the AVP but the study does not include it).
\end{enumerate}

Additionally, of the 218 records identified, 2 were duplicate publications of the same study and were removed before screening, leaving 216 records for the first screening, in which the first screener applied the criteria above. 
That stage excluded 106 records (76 not primarily about AVP, 24 by document type, and 6 with no usable abstract) and retained 110. 
In the second screening, a second reviewer checked all of the first screener's decisions and, in a separate blind pass on the same updated export, screened every record without sight of those decisions. 

The second reviewer flagged eight retained papers in which AVP might not be the central subject, and the first screener reread them against the criteria.
Six were excluded at this stage: one essay on smart glasses in general, two systems built for multiple headsets, and three studies that used AVP only as a research instrument. 
The other two \cite{12, 105} were retained as judgment calls under criterion~2. 

Each measures AVP's performance against a no-headset baseline and against one optical-see-through headset. AVP is the device under evaluation in both, but the presence of a second measured headset places them at the boundary of that criterion. 
Because the eight records were resolved through discussion, their reclassification is a consensus decision rather than an independent one. 
Although the blind pass would permit one, we did not compute an inter-rater agreement statistic, which we note as a limitation in Section~\ref{sec:discussion}.

The two stages together excluded 112 records (82 not primarily about AVP, 24 due to document type, and 6 with no usable abstract), leaving 104 papers included.

\Cref{fig:prisma} summarizes this PRISMA-informed screening flow \cite{113}, and we return to the composition of the resulting corpus below (\Cref{sec:discussion}) as a finding in its own right about the state of the field. 
This 104-paper corpus is supplemented by 14 further references, cited throughout the text, that support methodological or contextual claims (instrument and model definitions, manufacturer specifications, a security advisory, and pricing sources) rather than being part of it.

\begin{center}
\figurehere
\includegraphics[width=\linewidth]{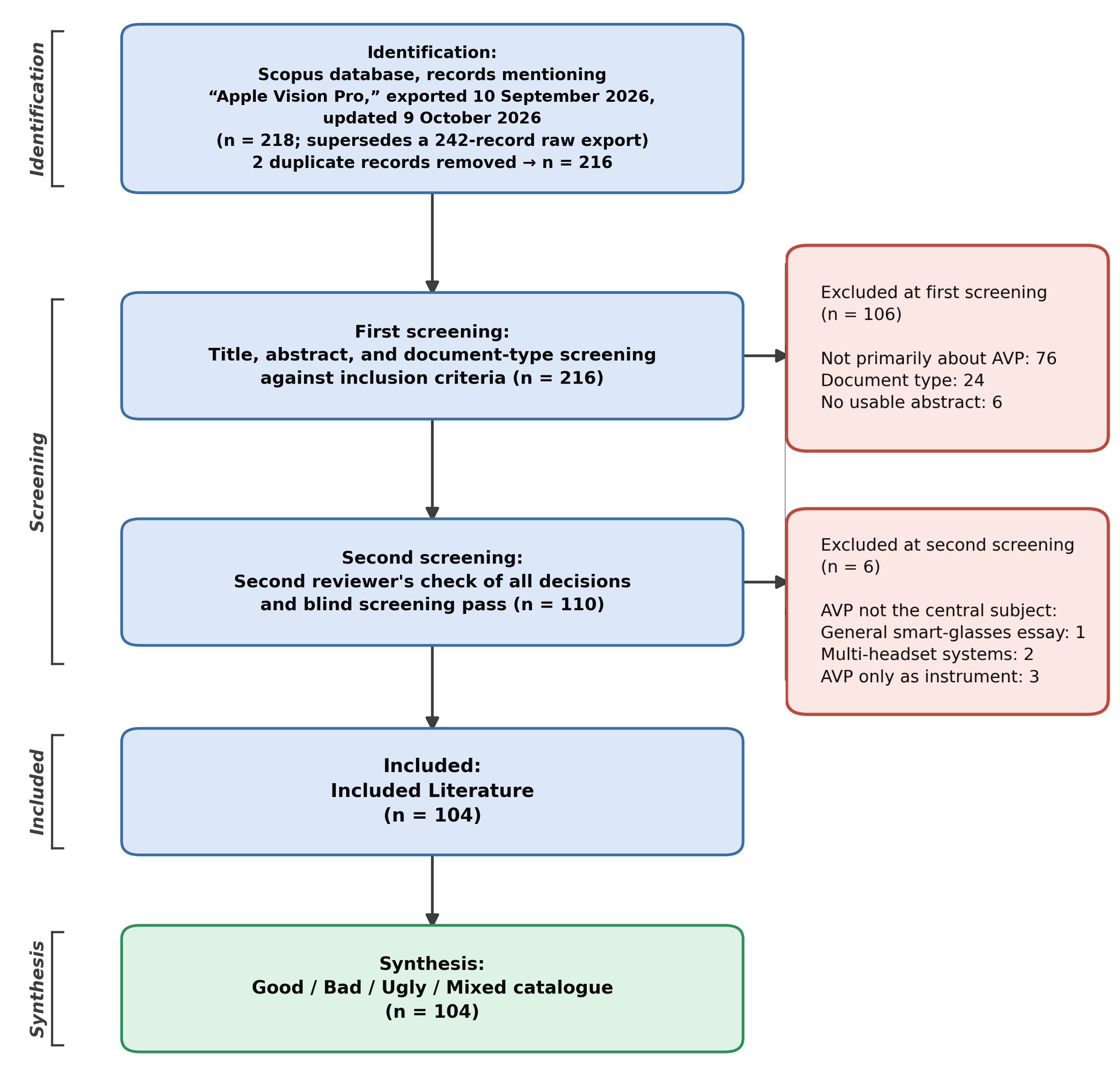}
\caption{Screening flow for the Scopus export underlying this review, from the initial search (n = 218), through duplicate removal (n - 2 = 216) and first-screening exclusion (n - 106 = 110) and second-screening exclusion (n - 6 = 104), to the included corpus analyzed.}
\label{fig:prisma}
\end{center}

\subsection{Analytical framework}\label{sec:framework}

We coded every included paper's primary contribution into exactly one of four buckets, operationalizing the tripartite structure introduced above for this specific device rather than for head-mounted displays in general.

This good/bad/ugly/mixed coding reflects a paper's \textit{stated primary contribution}, not its statistical outcome: a study whose central aim was to validate a benefit was coded good even when its own results were null or mixed on that benefit (e.g. \cite{107} in \Cref{sec:good}), provided the paper's framing, design, and discussion remained oriented toward that validating aim. 
We flagged in \Cref{sec:discussion} that this rule likely biases the good count upward relative to a coding scheme keyed to the statistical outcome rather than the stated aim.

\subsubsection{A ``good'' classification}

A paper was coded \textit{good} when its central contribution was a validated or working application: a feasibility study, usability evaluation, or system description reporting that AVP accomplished a stated task.

\subsubsection{A ``bad'' classification}

A paper was coded \textit{bad} when its central contribution documented a technical, ergonomic, or software-ecosystem limitation, a constraint the authors themselves frame as a target for future engineering rather than as a deliberate design or deployment choice.

\subsubsection{An ``ugly'' classification}

A paper was coded \textit{ugly} when its central contribution concerned an ethical, social, or structural consequence of a choice already made in how AVP is built, marketed, priced, or deployed, a consequence that engineering iteration alone would not resolve.

\subsubsection{A ``mixed'' classification}

A fourth outcome, \textit{mixed}, was recorded when a paper's own reported findings substantively supported more than one of these registers rather than a single one dominating; such papers are cited in every section to which their evidence speaks; \Cref{sec:discussion} discusses several concrete examples.

This coding produced 68 good, 7 bad, 2 ugly, and 27 mixed papers (68 + 7 + 2 + 27 = 104); \Cref{tab:coding} summarizes the split. 
Reference numbers are assigned in first-citation order in the text, not by bucket, so they do not align directly with this split. 
A mixed paper's first citation typically falls in whichever section happens to introduce it, even though the same paper is also cited later in one or two other sections.

\begin{table}[pos=H]
\centering
\caption{Good/bad/ugly/mixed coding of the 104 included papers, with representative references.}
\label{tab:coding}
\small
\begin{tabular}{ccc}
\toprule
\textbf{Bucket} & \textbf{Count} & \textbf{Representative refs} \\
\midrule
Good  & 68 & \cite{2,9,27} \\
\addlinespace
Bad   & 7  & \cite{90,91,93,96} \\
\addlinespace
Ugly  & 2  & \cite{99,103} \\
\addlinespace
Mixed & 27 & \cite{42,49,60,73} \\
\bottomrule
\end{tabular}
\end{table}

\Cref{fig:growth} characterizes the resulting corpus by publication year: from 11 primary-focus papers in the device's first partial calendar year of availability (2024) to 47 in 2025, with 46 more already recorded through 9 October 2026, a trajectory consistent with a research community still in the early, accelerating phase of studying a new artifact.
\pagebreak

\begin{center}
\figurehere
\includegraphics[width=\linewidth]{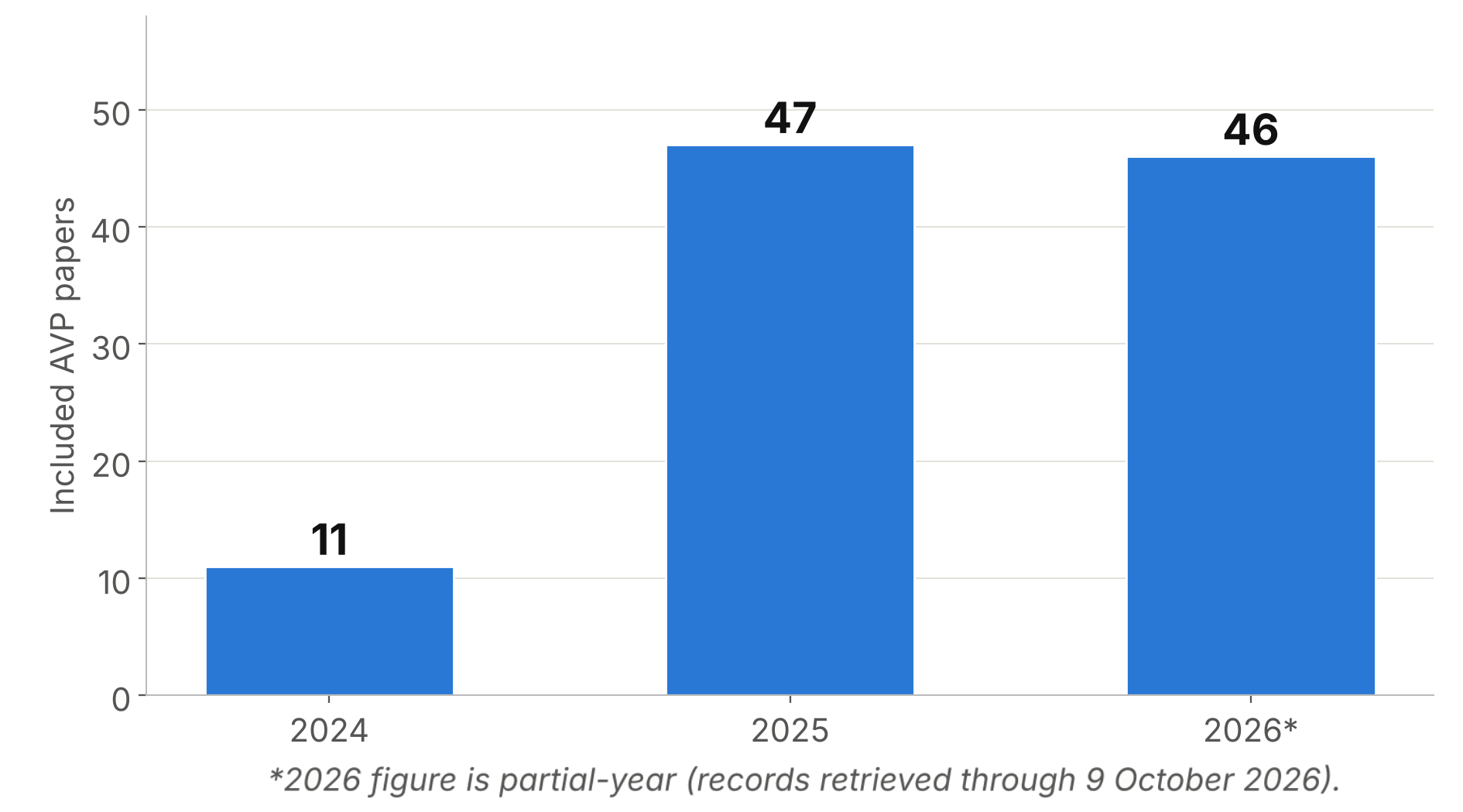}
\caption{Growth of primary-focus Apple Vision Pro research in the included corpus (n = 104), by publication year. The 2026 figure is partial-year, covering records retrieved through 9 October 2026.}
\label{fig:growth}
\end{center}

Alongside this good/bad/ugly coding, we classified each included paper by its primary application domain (medicine and surgery, education, or software tooling) to characterize the composition of the corpus independently of the tripartite catalog and to structure the sub-themes reported within \Cref{sec:good,sec:bad,sec:ugly}. \Cref{fig:donut} shows the resulting distribution: medicine and surgery, education and training, and the software and developer ecosystem together account for over half of the 104 included papers, a concentration we return to when discussing where the evidence base most needs to grow.

\begin{center}
\figurehere
\includegraphics[width=\linewidth]{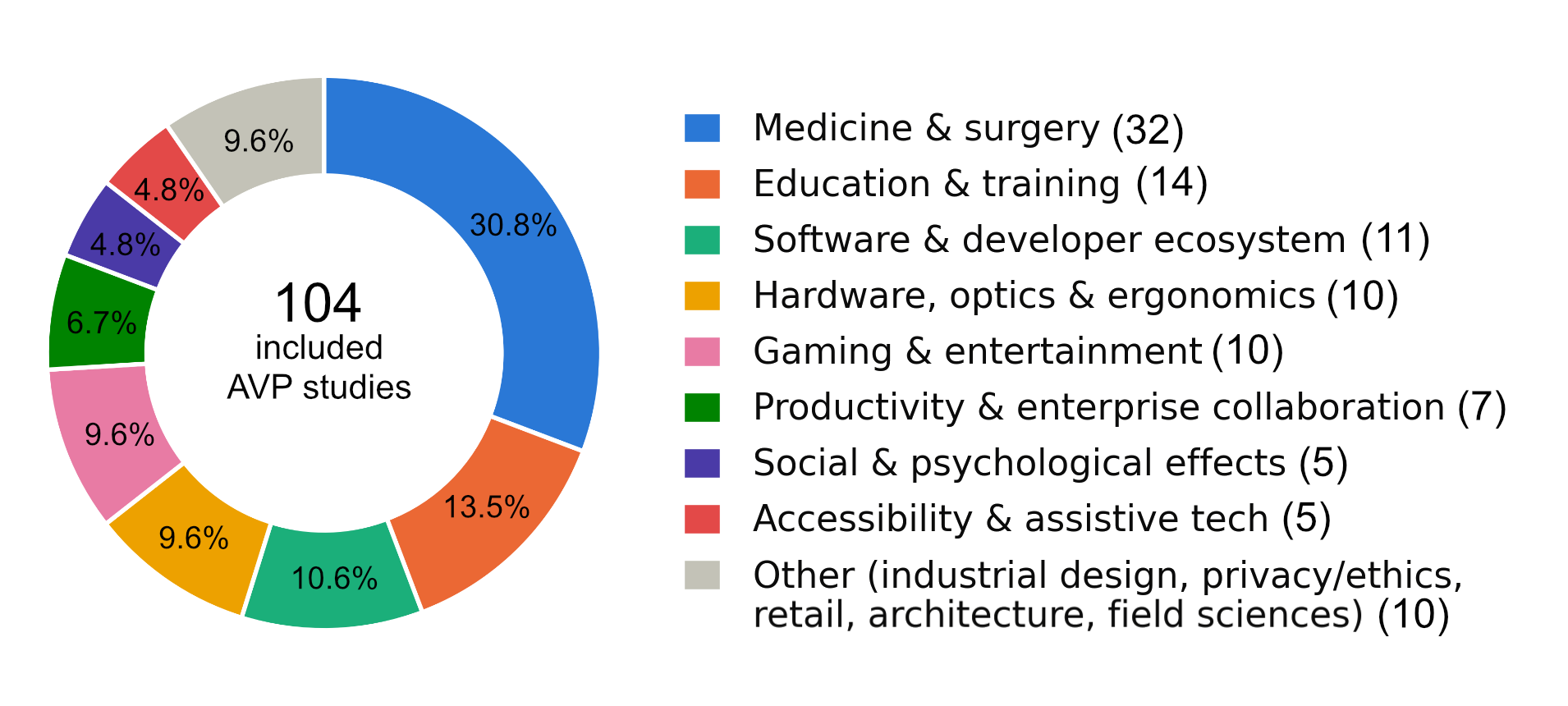}
\caption{Application domains of the 104 papers included in this review (2024--2026).}
\label{fig:donut}
\end{center}

Because almost every included paper's data collection predates the October 2025 M5 refresh, this corpus overwhelmingly reflects the original M2-based hardware; the ergonomic and performance findings cataloged in \Cref{sec:bad} in particular should not be assumed to transfer unchanged to the M5 model, since a heavier, differently balanced headset plausibly shifts exactly the comfort and workload outcomes that literature measures: the M5 model weighs 750--800g with its redesigned, counterbalanced Dual Knit Band, versus 600--650g for the original model's Solo Knit Band and Dual Loop Band options \cite{121,122}, and it supports a higher passthrough refresh rate.

\pagebreak

\section{The good}\label{sec:good}

The literature converges on a clear pattern: wherever a task benefits from precise spatial registration, embodied interaction, or head-worn stereoscopic visualization, the AVP has been shown to work, often on the first attempt. 
The following sub-themes summarize where the corpus reports validated, positive, or genuinely promising outcomes.

\subsection{Surgical visualization and intraoperative guidance.} 
The single largest and most consistent cluster of positive findings concerns intraoperative use of the AVP as a wearable, wireless display that consolidates multiple imaging feeds directly into the surgeon's line of sight. 
In ophthalmology, prospective feasibility studies report high procedure-completion rates and reduced surgeon workload when the headset served as the primary visualization platform for cataract surgery, endoscopic dacryocystorhinostomy, intraocular surgery with live telementoring, and oculoplastic procedures, with usability ratings consistently in the ``excellent'' range \cite{2,3,4,5}. 

Hepatobiliary and abdominal surgeons report similar gains: an AVP-based real-time image-fusion navigation system increased puncture accuracy from 42.5\% to 90\% in a comparative experiment \cite{6}, separate from the same study's eight-patient clinical series, which was completed without major complications. 
augmented reality (AR)-assisted cholangioscopy and laparoscopic radiofrequency ablation were completed without technical failure or complications \cite{7,8}, and the device played a documented role in guiding lesion identification during the first Da Vinci-assisted total gastrectomy performed with its aid \cite{9}. 

In spine and orthopedic procedures, AR-guided plate positioning and biportal or standard endoscopic lumbar surgery were completed successfully with head-mounted visualization replacing or supplementing conventional monitors, without added cognitive workload or complications \cite{10,11,12}. 
Minimally invasive general surgery, urologic Mini-ECIRS, and chest-wall oncologic reconstruction case series likewise report low perceived workload, uneventful outcomes, and workflow compatibility \cite{13,14,15}, while a co-designed multimodal assistant demonstrates that sterile, glanceable AR interaction can be built around interventional radiology's specific procedural constraints \cite{16}. 

A scan-derived digital-twin framework further shows sub-2mm registration accuracy for surgical robotic components rendered on the device \cite{17}, and immersive endoscopy reduces neck and shoulder muscle activation relative to a non-ergonomic monitor arrangement \cite{18}. A neurosurgical planning application built natively for the headset, with controller-free manipulation of patient-specific 3D models and synchronized multi-user case discussion, received a total satisfaction score of $+1.7$ on a scale from $-2$ to $+2$ from 15 neurosurgeons \cite{123}.

\subsection{Surgical, procedural, and clinical training.} 
Beyond live intervention, AVP-based simulators and stepwise AR overlays show measurable learning gains across several specialties. 
A stepwise AR learning-support system for surgical education and an AVP-integrated laparoscopic surgery teaching program each report improved trainee performance, with the latter documenting a 15\% improvement in operative skill and a 9\% increase in exam scores after iterative latency optimization \cite{19,20}. 
Domain-specific MR trainers outperform lecture- or screen-based instruction: an ultrasound-interpretation tool for veterinary sonography improved both immediate comprehension and six-week retention relative to lecture-trained peers \cite{21}, a naloxone-administration trainer increased knowledge, confidence, and intent to help across a heterogeneous lay and professional sample \cite{22}, and a resident vascular-surgery case study found superior spatial awareness of a 3D aortic model compared with standard workstation viewing \cite{23}. 

A hybrid femoral-nailing simulator combining real instruments with AR guidance produced clear procedural-efficiency gains that were retained at follow-up \cite{24}. At the same time, quantitative eye-tracking benchmarking supports the device's suitability for training cardiac surgeons in fine-grained anatomical landmark selection \cite{25}.

\subsection{Diagnostic imaging, radiology, and pathology workflows.} 
Outside the operating room, pilot deployments in radiology and pathology point to strong short-term feasibility. 
A prospective pilot with 25 radiology physicians (22 residents, 2 fellows, 1 attending) found rapid onboarding (under five minutes), high perceived utility for multidisciplinary conferences, and majority-positive ratings of ease of use \cite{26}. 
Cinematic rendering software for volumetric computed tomography (CT) and magnetic resonance imaging (MRI) review was rated highly usable by 14 medical experts on the platform \cite{27}, and a gigapixel whole-slide pathology viewer that combines embodied navigation with multimodal AI assistance demonstrated technical feasibility for reducing the panning and context-switching burden of conventional 2D digital pathology \cite{28}. 

In interventional settings, AR and VR preprocedural planning changed the surgical approach in nearly half of 50 hepatic oncology cases and yielded high scores in anatomic understanding and procedural confidence \cite{29}. 
At the same time, comparative phantom studies of AR-guided percutaneous biopsy and ultrasound-guided needle placement found accuracy and workflow benefits comparable to or exceeding those of conventional guidance, with novice operators gaining the most \cite{30,31}. 

A pilot comparing headset- and desktop-based interpretation of diverticulitis on CT found statistically equivalent diagnostic accuracy between the two \cite{32}. Elsewhere, an AR-guided functional near-infrared spectroscopy (fNIRS) probe-positioning system achieved sub-centimeter, hands-free placement accuracy for neurorehabilitation applications \cite{33}, and a review outlines the device's potential as a portable, home-based perimetry tool for glaucoma monitoring \cite{34}.

\subsection{Education and embodied cultural learning.} 
A recurring finding across the education literature is that embodied, spatially situated interaction produces measurable learning gains over flat or GUI-based alternatives. 
A controlled study of a Chinese-artifact learning platform found significantly higher learning gains, engagement, and interest than a traditional GUI equivalent \cite{35}, and a spatial-computing chair-design curriculum produced more intuitive, creative design outcomes than an iPad-based control condition \cite{36}. 

Museum and heritage applications extend this pattern: an MR system for exploring a national museum's storage and curatorial practices \cite{37}, an eXtended reality (XR) documentary format developed explicitly in response to the device's release for visible-storage exhibitions \cite{38}, and a retrieval-augmented, voice-and-gesture interface for conversing with digitized historical archives \cite{39} all demonstrate novel access to cultural content that flat interfaces do not afford. 

A reproducible cardiovascular-anatomy platform paired with a retrieval-augmented voice tutor met its real-time interaction targets across eight anatomical systems while substantially undercutting the licensing cost of proprietary alternatives \cite{40}. 
Field-based and blended-learning studies add nuance without displacing the overall positive signal: mixed-reality note-taking during outdoor excursions improved content extraction for physical and textual information relative to pen-and-paper and mobile notes \cite{41}, and a comparison of classical media against XR eLearning found immersive presentation boosted engagement with visually rich content, with most participants keen to use XR for learning again \cite{42}. 

Adoption-modeling and controlled-comparison research adds a further, more mixed data point: a study of AVP's AI-assisted features using PLS-SEM, partial least squares structural equation modeling, found most theorized predictors of continued use confirmed, though informativeness gratification did not significantly predict use \cite{106,116}, and a counterbalanced comparison of Persona-mediated mixed-reality vocal instruction against tablet-based video and in-person teaching found only a non-significant trend toward reduced performance anxiety, with no significant difference in sight-singing outcomes overall \cite{107}.

\subsection{Accessibility and assistive applications.} 
Several teams treat the device's depth-sensing, tracking, and passthrough capabilities as infrastructure for accessibility research rather than as an end in themselves. 
A technical feasibility study used AVP's depth detection and semantic segmentation to identify trip hazards exceeding 1.5 cm in height, a step toward real-time fall-prevention support \cite{43}, and a companion VR study used the platform to build ecologically realistic obstacle scenarios for studying older adults' walking safety, isolating an interaction effect between static and dynamic hazards \cite{44}. 

A co-design study with adults with Cerebral Visual Impairment used the headset across a two-week diary study and iterative development sessions to address object location, text reading, face recognition, and sensory stress management in a population underrepresented in assistive technology research \cite{45}. 
An AR interface built on the device's tracking and object-detection capabilities to support semi-autonomous prosthetic-hand control significantly lowered workload and raised usability scores for participants open to wearing a headset \cite{46}. An in-the-wild evaluation with Deaf and hard-of-hearing users likewise revealed concrete opportunities to make spatial computing more inclusive for this population \cite{47}.

\subsection{Design, architecture, and industrial engineering workflows.} 
In design and engineering contexts, immersive 1:1-scale evaluation is repeatedly credited with judgments that flat screens do not support. 
An architectural design study found that while tablet interaction accelerated idea generation, spatial immersion on the headset enabled a deeper perceptual understanding of mass, light, and enclosure, shaping design choices beyond what screen-based tools afford \cite{48}. A biomimetic transportation design study similarly found that immersive scale and depth perception enhanced the assessment of proportions during AI-assisted vehicle styling \cite{49}. 

In structural engineering, an interactive topology-optimization playground built natively for the device integrates real-time human judgment into compliance-minimization workflows, aiming to reduce lead times and improve manufacturability \cite{50}. A natively redesigned interface for teleoperating an underwater remotely operated vehicle achieved the highest usability scores and the lowest workload among the three tested interfaces, outperforming both a tablet-based control and a direct port of that interface \cite{51}. 

Finally, an immersive digital supply chain twin prototype was evaluated as a step toward more transparent, collaborative planning among network partners \cite{52}.

\subsection{Spatial video, novel sensing, and hardware-enabled research instrumentation.} 
The device's built-in stereoscopic capture, precise eye- and hand-tracking, and depth sensing have made it a research instrument in its own right. 
Public datasets built around its native spatial-video pipeline, including a 300-clip stereoscopic dataset, a quality-of-experience dataset showing that spatial video delivers superior perceptual quality at matched bitrates, and a 600-clip egocentric spatial-video quality-assessment database, address a previously missing resource for codec and immersive-video research \cite{53,54,55}. 

Outside the lab, the platform's depth sensing supported forestry-inventory apps whose tree-diameter measurements closely tracked caliper references (R² up to 0.978) across varied terrain \cite{57}, and custom haptic gloves with vibrotactile and thermoelectric feedback were built around its hand-tracking to extend tactile realism in AR \cite{58}. 
Vision scientists further show that classic psychophysical phenomena, such as visual crowding, can be replicated on consumer headsets for trial designs in which continuous gaze monitoring is not required \cite{59}. 

A click-based gaze-visualization tool demonstrates that useful attention heatmaps can still be extracted for research purposes within the platform's constraints on continuous access to eye-tracking data \cite{60}.

\subsection{Gaming, entertainment, and immersive art and storytelling.} 
Creative and entertainment applications exploit the headset's passthrough, spatial audio, and unbounded interaction volume. 
Comparative interaction studies find the device's gaze-pinch and direct-touch input modalities functionally equivalent for rapid target acquisition in a custom gaming task, giving developers latitude in interaction design \cite{61}. 
A real-time hand-tracking boxing trainer achieved 96.3\% gesture-recognition accuracy at 60 FPS across lighting conditions, extending precision sports coaching beyond gym settings \cite{62}. 

Physics-driven interaction demonstrations, a dart-throwing simulator rated highly for realism and immersiveness, and a block-stacking task probing depth perception and kinetic control, show that native hand- and eye-tracking support convincing unbounded-volume physical simulation \cite{63,64}. 
Narrative and installation work extends the medium further: a validated ``Point-Line'' framework for MR cinema significantly improved narrative efficacy and lowered cognitive load relative to conventional immersive video \cite{65}, particle-based video-content extension into the surrounding room enhanced presence and concentration relative to a character-based baseline \cite{66}, and standalone art installations exploring embodied perception \cite{67} and processed digital experience \cite{68} illustrate the platform's use as a venue for spatial art. 

An exclusively built 3D multiplayer game \cite{70} and a publicly released, theremin-inspired polyphonic instrument controlled by hand tracking \cite{124} round out the entertainment cluster.

\subsection{Productivity, collaboration, and natural spatial interaction.} 
Workplace-oriented studies find that natural, gesture- and gaze-based interaction is readily and quickly adopted. Participants with little prior XR exposure engaged with 3D content on the headset without instruction, favoring embodied gesture and voice input over keyboard-based alternatives \cite{71}. 
Systematic manipulation of immersion level found that selective attention improved as immersion increased during individual work \cite{72}, and interviews with ten long-term users found that self-reported productivity increased and users adapted to the device over time \cite{73}. 

FaceTime's spatial-persona telepresence feature was found to deliver a genuinely immersive experience at lower bandwidth than competing 2D-persona video conferencing \cite{74}, and a business-intelligence dashboard ported to visionOS was received with enthusiasm for its potential to promote new engagement patterns with enterprise data at a large industry conference \cite{75}. 

An AR task-guidance framework combining voice input, large language models, and object detection likewise enhanced participant engagement during a real-world guided task \cite{76}.

\subsection{Developer tooling and generative spatial-content creation.} 
A cluster of platform-specific toolkits and creation apps shows how quickly a native content-creation ecosystem formed around the headset's spatial APIs. 
Toolkits such as Palimpsest automate the coherent spatial arrangement of 3D content around tracked real-world objects \cite{77}, MediaMix built a full multimedia-retrieval interface around the device's eye-tracking and gesture recognition for a video-retrieval competition \cite{78}, and RécitKit demonstrated that spatial data narratives, illustrated with an immersive historical flow map, support insight formation for both authors and viewers \cite{79}. 

Generative-AI creation tools, including AI3D Desktop and VolumeMatic, allow text- or voice-driven 3D content generation directly in the headset \cite{80,81}, and a related system, Dream Space, found in a 20-participant mixed-methods evaluation that this spatial-computing creation paradigm significantly reduced cognitive load relative to conventional 3D tools and let novices complete complex creative tasks within 20–30 minutes \cite{82}. 

Immersive analytics applications extend this tooling to technical debugging, with computer vision and machine learning (CVML) engineers reporting that seamlessly integrating 2D and 3D visualization on the device improved their evaluation workflow for a spatial localization model \cite{83}, while a scene-aware spatial-measurement prototype \cite{84} and an XR sketch-to-3D-generation system \cite{86} further illustrate the breadth of native tools now built for the platform.

\subsection{Memory and mixed-reality content.} 
Beyond task performance, one study examines how AVP-mediated content is remembered. 
A passthrough object-viewing study identified the specific perceptual cues, namely lighting, shadow, texture, and tactile exploration, that shape how well real and virtual objects are later remembered, informing design for memorable mixed-reality content \cite{88}. \Cref{tab:good} summarizes these sub-themes.

\pagebreak

\begin{strip}
\centering
\tablehere
\caption{The good: application areas where AVP has been validated, mostly in feasibility and pilot studies.}
\label{tab:good}
\small
\begin{tabular}{p{0.18\linewidth}p{0.6\linewidth}p{0.14\linewidth}}
\toprule
\textbf{Application area} & \textbf{Main findings} & \textbf{Refs} \\
\midrule
Surgical visualization and intraoperative guidance & Wearable, wireless intraoperative display consolidating multiple feeds; high completion rates, low workload, improved navigation accuracy across ophthalmology, hepatobiliary, spine, general, urologic, and interventional-radiology procedures & \cite{2,3,4,5,6,7,8,9,10,11,12,13,14,15,16,17,18,123} \\
\addlinespace
Surgical, procedural, and clinical training & Stepwise AR overlays and MR simulators outperform lecture- or screen-based training in skill acquisition, exam scores, retention, and confidence & \cite{19,20,21,22,23,24,25} \\
\addlinespace
Diagnostic imaging, radiology, and pathology workflows & Rapid onboarding and strong perceived usability for image review, cinematic rendering, whole-slide pathology, and preprocedural AR/VR planning, with accuracy comparable to conventional tools & \cite{26,27,28,29,30,31,32,33,34} \\
\addlinespace
Education and embodied cultural learning & Embodied, spatially situated interaction improves learning gains, engagement, and access to museum/heritage content over flat interfaces & \cite{35,36,37,38,39,40,41,42,106,107} \\
\addlinespace
Accessibility and assistive applications & Depth sensing, tracking, and passthrough repurposed for fall-hazard detection, low-vision support, prosthetic control, and Deaf/hard-of-hearing accessibility research & \cite{43,44,45,46,47} \\
\addlinespace
Design, architecture, and industrial engineering workflows & Immersive 1:1-scale evaluation improves spatial/proportion judgment and interface usability in architecture, transportation design, topology optimization, teleoperation, and supply-chain visualization & \cite{48,49,50,51,52} \\
\addlinespace
Spatial video, novel sensing, and hardware-enabled research instrumentation & Native stereoscopic capture, eye/hand tracking, and depth sensing enable new public datasets, forestry measurement, haptics, and portable psychophysics & \cite{53,54,55,57,58,59,60} \\
\addlinespace
Gaming, entertainment, and immersive art and storytelling & Passthrough, spatial audio, and unbounded interaction volume support functionally equivalent input schemes, precision sports training, physics-driven play, gestural musical instruments, and novel narrative/installation formats & \cite{61,62,63,64,65,66,67,68,70,124} \\
\addlinespace
Productivity, collaboration, and natural spatial interaction & Gesture- and gaze-based interaction is adopted quickly and without instruction; immersion improves selective attention; spatial telepresence and immersive analytics show early workplace promise & \cite{71,72,73,74,75,76} \\
\addlinespace
Developer tooling and generative spatial-content creation & A native toolkit and generative-AI content-creation ecosystem lowers cognitive load and technical barriers for 3D authoring, data narratives, and debugging & \cite{77,78,79,80,81,82,83,84,86} \\
\addlinespace
Memory and mixed-reality content & A passthrough object-viewing study clarifies which perceptual cues make mixed-reality objects memorable & \cite{88} \\
\bottomrule
\end{tabular}
\end{strip}

\section{The bad}\label{sec:bad}
The limitations documented in this section are distinct in kind from the concerns raised in \Cref{sec:ugly}. 
Where ``the ugly'' describes deliberate choices in how AVP is built, marketed, or used (choices that raise ethical questions about consent, surveillance, or exclusion), ``the bad'' describes the persistent technical and operational immaturity of a headset still in its first two hardware iterations. 

These are not intentional harms: they are the ergonomic, perceptual, and software-ecosystem constraints that recur across the controlled studies, feasibility trials, and field evaluations reviewed here, and that most of these same papers frame as targets for iteration rather than as design failures. 

Notably, only seven of the 104 included papers were coded as centrally organized around such limitations (with a further 11 of the 27 mixed-coded papers also touching on them as one of several findings, discussed as ``mixed'' in \Cref{sec:method}); far more report them as secondary findings inside otherwise favorable feasibility or usability studies, which is itself informative about how deeply these constraints are embedded in the current hardware generation regardless of application domain.

\subsection{Ergonomic burden and physical side effects.} 
Multiple independent evaluations converge on the same complaint: the headset's weight and bulk become a liability over time or in professional settings. A case report of endoscopic spine surgery notes that ``current AR headsets are relatively large and bulky,'' which ``may cause discomfort from the headset's weight if used for long periods of time'' \cite{12}. 
An exploratory study of young adults using AVP within visionOS found that, despite participants' high technology familiarity and screen-time exposure, they ``experienced physical side effects early in the testing process,'' prompting the authors to argue explicitly for more attention to negative physical outcomes alongside XR's celebrated benefits \cite{90}. 

Even in a largely positive feasibility pilot of AVP for diagnostic radiology, 3 of the same 25 participants (12\%) reported eye strain after a 14-minute average session \cite{26}. A mixed-methods evaluation combining structured user-experience feedback with technical assessment similarly reported measurable UX improvement over prior headsets alongside persistent comfort and ergonomic pain points that future hardware generations still need to address \cite{109}.

\subsection{Input-method trade-offs and interaction limits.} 
AVP's signature gaze-and-pinch interaction paradigm underperforms simpler alternatives on measurable tasks. 
A controlled comparison of AVP's two native text-entry methods found Tap Typing significantly faster and more accurate than Gaze + Pinch, with participants also preferring it and rating it higher on usability and workload measures \cite{91}. 

A study of automatically arranged multi-window layouts on AVP found no significant improvement over conventional arrangements and surfaced ``fundamental challenges common to multi-window tasks in HMD environments, such as difficulties with text input, visual feedback, and physical strain'' \cite{92}. 

Vision-science researchers attempting to repurpose AVP as a portable psychophysics instrument likewise report concrete platform constraints: the device provides no access to continuous real-time eye-position data, and visionOS specifies stimulus size and position in device-defined ``points'' rather than degrees of visual angle, complicating direct comparison with laboratory-grade eye trackers \cite{59}. A closely related workaround study built a click-based gaze-logging tool precisely because AVP's privacy architecture ``prevent[s] direct access to continuous user gaze data,'' forcing researchers toward manual pinch- or controller-triggered capture with substantially lower sampling rates than true gaze tracking \cite{60}. 

In a design-generation context, voice and freehand-sketch input on AVP were found to ``lack the precision of traditional desktop workflows,'' producing designer–AI intent mismatches during iterative 3D design tasks \cite{49}.

\subsection{Performance and workload costs relative to established alternatives.} 
Several controlled comparisons quantify a real performance penalty for AVP relative to both unencumbered vision and competing headsets. 
In a suturing-technique study with 20 healthcare professionals, AVP produced significantly longer task completion times, significantly higher workload on the NASA Task Load Index (NASA-TLX) \cite{117}, significantly lower system usability, and significantly higher simulator sickness scores than both Microsoft HoloLens 2 and a no-headset baseline \cite{93}. 

A companion study isolating video-see-through optics from optical-see-through and unencumbered vision found that manual dexterity, measured with the Purdue Pegboard Test \cite{118}, a timed test of fine motor speed using pins and washers, ``significantly declined under the AVP condition across all subtests,'' accompanied by higher cognitive load and post-task reports of discomfort and visual fatigue, a ``hidden ergonomic cost'' the authors argue designers of AR-assisted tools must actively counterbalance \cite{94}. 

In a phantom-based interventional radiology comparison, AVP was rated more mentally demanding than an optical-see-through competitor, particularly among experienced operators \cite{30}. 
In diagnostic radiology, the median time per case for detecting diverticulitis on CT was significantly longer on AVP (57 s) than on a conventional desktop workstation (31 s), despite equivalent diagnostic accuracy \cite{32}.

\subsection{Software and application ecosystem immaturity.} 
Beyond individual task performance, reviewers repeatedly note that the surrounding software ecosystem has not matured to match the hardware. 
Interviews with ten long-term AVP users conducted more than a year after release found that, in spite of marketing claims, persistent limitations, ``a lack of native applications, difficulties when integrating XR into current workflows, and limited possibilities to adapt and customize the XR experience,'' continue to constrain everyday productivity use \cite{73}. 

A graphics-visualization study attempting GPU-based volume rendering on visionOS found the platform's native APIs ``predominantly oriented towards surface shading,'' requiring workarounds for a technique common in medical and manufacturing applications \cite{95}. 

A domain review in archaeology concluded, after direct field testing, that AVP ``does not yet represent a significant enough shift forward to have a major impact'' on the discipline compared to prior XR hardware \cite{96}. 
Pedagogically, a comparison of classical media against AVP-based eLearning found a clear preference for traditional formats for reading and writing tasks, with knowledge retention measurably lower for text-heavy material presented in the immersive environment \cite{42} (the same comparison reported higher engagement for visually rich content in \Cref{sec:good}, one study cutting both ways depending on content type, an example of the single-bucket-coding tension discussed in \Cref{sec:discussion}.)

\subsection{Connectivity, latency, and the cost of small-sample validation.} 
Streaming and power constraints round out the picture. A position paper on spatial video streaming to AVP identifies ``bandwidth limitations, latency, and synchronization issues between multiple camera views'' as unresolved obstacles to viable spatial-video delivery \cite{97}. 
A medical-education deployment for laparoscopic surgery teaching reported that initial video latency of 120 ms had to be reduced through resolution trade-offs and hardware upgrades before the system was considered clinically usable, ultimately reaching under 60 ms only after additional investment in capture cards and workstations \cite{20}. 

On power, an early clinical-adoption survey notes that although AVP is ``untethered,'' it remains tied to the user via ``the cable to the battery,'' a compromise this device shares with, but does not fully escape from, earlier tethered headsets \cite{98}. 

Cost recurs as a stated constraint on study design rather than an equity argument: a hepatic-ablation navigation study lists ``cost and fatigue'' alongside its positive accuracy results as open challenges to broader deployment \cite{6}, and a volume-rendering evaluation explicitly frames AVP's ``initial high cost'' as a present barrier to wider platform access for developers and researchers, one it expects to erode only as the device matures in later hardware generations \cite{95}. 
Consistent with this, the great majority of clinical AVP evaluations surveyed for this review remain single-case reports or small (single-digit-to-low-double-digit) pilot cohorts, a pattern consistent with, though not conclusively caused by, the device's price, limiting the sample sizes and statistical power that early adopters can field. 

Consumer adoption research offers a complementary, demand-side view rather than additional evidence on cost. A technology acceptance model survey of 272 sport fans found that purchase intention toward AVP was shaped indirectly, through perceived usefulness and ease of use, by team identification with the sport in question \cite{108,115}. Like the small-sample clinical literature, this work treats willingness to buy, rather than clinical or educational outcome, as its dependent variable.

Taken together, these findings describe a device still in active technical adolescence: capable enough to enter operating rooms, classrooms, and laboratories, but not yet comfortable, precise, well-supported, or affordable enough to do so at scale without measurable costs to its users. \Cref{tab:bad} summarizes these limitations.

\begin{strip}
\centering
\tablehere
\caption{The bad: recurring technical and ergonomic limitations of AVP, and their main findings.}
\label{tab:bad}
\small
\begin{tabular}{p{0.18\linewidth}p{0.6\linewidth}p{0.14\linewidth}}
\toprule
\textbf{Limitation} & \textbf{Main findings} & \textbf{Refs} \\
\midrule
Ergonomic burden and physical side effects & Headset weight/bulk causes discomfort over extended use; early physical side effects reported in young, tech-familiar users; measurable eye strain even in short sessions. & \cite{12,90,26,109} \\
\addlinespace
Input-method trade-offs and interaction limits & Native Gaze + Pinch input is slower and less accurate than Tap Typing; multi-window text entry and visual feedback remain difficult; restricted/no continuous gaze-tracking API complicates research and precision use; voice/sketch input lacks desktop-level precision. & \cite{91,92,59,60,49} \\
\addlinespace
Performance and workload costs relative to established alternatives & Controlled studies show AVP produces slower task completion, higher cognitive workload, lower usability, higher simulator sickness, and measurable manual-dexterity decline compared to unencumbered vision or competing headsets (e.g., HoloLens 2, desktop workstations). & \cite{93,94,30,32} \\
\addlinespace
Software and application ecosystem immaturity & Long-term users report a lack of native apps and workflow integration; visionOS graphics APIs are not well suited to some technical tasks (e.g., volume rendering); domain reviews find the platform not yet transformative; text-heavy eLearning content shows reduced retention in XR. & \cite{73,95,96,42} \\
\addlinespace
Connectivity, latency, and the cost of small-sample validation & Spatial-video streaming faces bandwidth, latency, and synchronization challenges; clinical deployments require substantial latency-engineering effort; the device remains tethered to an external battery; high cost is repeatedly cited as a barrier and coincides with a corpus dominated by small pilot studies and single-case reports. & \cite{97,20,98,6,95,108} \\
\bottomrule
\end{tabular}
\end{strip}

\section{The ugly}\label{sec:ugly}
Where the previous section cataloged limitations Apple and third-party developers are still working to solve, the ugly concerns a different register of problem: harms arising not from what the AVP cannot yet do well, but from what it already does, and from the choices developers, marketers, employers, and bystanders make around it. 

These are not bugs awaiting a firmware update; they are foreseeable consequences of design and deployment decisions: a camera array that is active whenever the device is worn, biometric-grade eye tracking, a photorealistic avatar system, and a launch price that sorts who gets to use, study, or be represented in this technology at all, combined with the promotional incentives of a nascent, heavily hyped consumer category. 

The literature is markedly thinner here than on the good or the bad: of 104 primary-focus AVP papers surveyed, only two were coded as centrally organized around misuse, risk, or exclusion (a further 6 of the 27 mixed-coded papers also touch on these concerns without them being the paper's stated primary contribution), and fewer still document realized rather than anticipated harm. That imbalance is itself part of the story, taken up explicitly below.

\textit{Gaze as an attack surface.} 
Apple markets AVP's eye-tracking-driven gaze-and-pinch input as a privacy-preserving improvement over hand- or acoustic-based interaction, since raw gaze coordinates are not exposed to third-party apps. 
GAZEploit shows this protection can be circumvented: by analyzing eye movements rendered onto a user's virtual Persona avatar during video calls or screen sharing, an attacker can remotely reconstruct what a wearer is typing on a gaze-controlled virtual keyboard, and the authors identify more than 15 top-rated App Store applications vulnerable to the technique \cite{99}. 

The feature meant to make the avatar feel present and expressive, its rendering of the wearer's eyes for others to see, becomes the side channel that leaks credentials and private text. Apple patched the specific vulnerability that GAZEploit exploited in visionOS 1.3, released on 29 July 2024 (CVE-2024-40865, affected component: Presence), by suspending Persona rendering while the virtual keyboard is active \cite{110}. 

The fix closes the one-channel GAZEploit demonstrated; it does not resolve the broader architectural tension the finding exposes, since AVP's privacy model still withholds continuous gaze data from third-party developers while rendering the wearer's eye movement outward through Persona in every other context. 
That a narrow, reactive patch was needed at all, and that it took independent academic security research, not an internal audit, to surface the exposure, is itself evidence for the \Cref{sec:discussion} point that this kind of finding is currently rare precisely because so little independent security research on AVP exists. 

Researchers building iTrace, a click-based gaze-visualization tool, show that even without continuous gaze access, user-triggered gaze samples (captured with a pinch or game-controller click) are sufficient to reconstruct attention heatmaps. 
They restrict their tool to research use in acknowledgment of how sensitive the recovered data is \cite{60}. 

Early-adopter reviews echo the concern: privacy is one of the few affordances adopters describe as actively discouraging rather than encouraging adoption \cite{100}.

\textit{Simulated presence, real absence.} AVP's Persona avatars and its EyeSight display, which projects a digital rendering of the wearer's eyes outward so bystanders can see where they are looking, are designed to preserve social presence while the wearer is immersed. 
A controlled comparison shows that the feature improves perceived social acceptability and presence relative to a headset without it, yet participants still rated it as falling short of unmediated face-to-face interaction \cite{102}. 

Read as a societal-harm question rather than a usability one, this is a notable admission: a technology marketed as compatible with ordinary social connection instead substitutes a simulacrum of eye contact for the real thing, and does so most visibly in exactly the shared, co-present situations where genuine attention matters most.

\subsection{Hype rhetoric and the suppression of critical scrutiny.} 
A recurring finding is that discourse around AVP has itself shaped how critically the device gets evaluated. 
A rhetorical analysis of YouTube reviews finds that anticipatory communities around the launch converged on a small set of strategies, namely excitement, exploration, exemplification, and exaggeration, that intensified anticipation while constraining more critical evaluation once the device shipped \cite{103}. 
Long-term users interviewed more than a year after launch push back against this rhetoric directly, describing persistent gaps between marketing's productivity promises and what everyday use actually delivers \cite{73}. 

Sentiment analysis of scraped online commentary likewise reports ``substantial variability'' in real-world reception rather than the consensus enthusiasm implied by early coverage \cite{104}. None of this is a technical failing of the headset; it is a pattern of promotional communication that a premium, still-young category is especially prone to, and it matters ethically because it delays the public's ability to reason clearly about the device's risks.

\subsection{Price as exclusion, not just a barrier.} 
The bad section treats AVP's price as a technology-adoption obstacle; read differently, the same fact is a structural equity problem. 
Market-behavior research applying the AIDA model \cite{119}, a marketing framework tracking attention, interest, desire, and action toward a purchase, finds purchase intentions above 75\% among high-income users, while price sensitivity substantially limits adoption among middle-income users \cite{105}, meaning the population able to buy, use, and be represented in AVP research is itself skewed toward the affluent. 

Because most of the ``good'' evidence in this corpus derives from small convenience samples of people who already own or can access an AVP, price does not merely slow diffusion; it shapes whose needs the emerging literature is even in a position to study.

\subsection{A conspicuously thin evidence base.} 
Finally, the shape of the corpus is itself a finding. 
Only two of the 104 included papers were coded as primarily ugly in focus, compared with 68 coded as good; most published AVP research to date consists of feasibility and pilot studies validating benefits in medicine, education, and design, rather than investigations of risk, misuse, or exclusion. We found no dedicated study in this corpus of, for example, employer or institutional surveillance uses of the headset, despite seven papers examining productivity and enterprise deployment, a gap worth naming rather than papering over with speculation.

Several of the concerns above resist being cleanly filed under either "bad" or "ugly". Price can be read as a temporary technical-adoption barrier (bad) or as a structural exclusion that determines who gets to benefit from and be represented in AVP research (ugly), depending solely on whether the observer's lens is the manufacturer's cost curve or the resulting research population. Diminished social presence is similarly double-faced: a limitation of current avatar and passthrough fidelity (bad) and, at the same time, a societal-harm consequence of choosing to mediate embodied co-presence through a commercial device at all (ugly). 

The gaze-tracking privacy architecture is a third such case: withholding continuous gaze data from developers while rendering eye movement outward through Persona can be read as an unfinished engineering compromise (bad) or as a considered trade-off favoring avatar expressiveness over wearer privacy (ugly). We do not resolve these ambiguities here; we return to them, without resolving them, in Section~\Cref{sec:discussion}.
\Cref{tab:ugly} summarizes these issues.

\begin{strip}
\centering
\tablehere
\caption{The ugly: ethical concerns and structural harms arising from how AVP is built, priced, and deployed.}
\label{tab:ugly}
\small
\begin{tabular}{p{0.18\linewidth}p{0.6\linewidth}p{0.14\linewidth}}
\toprule
\textbf{Concern} & \textbf{Main findings} & \textbf{Refs} \\
\midrule
Gaze as an attack surface & Eye-tracking and Persona-avatar rendering intended to improve privacy and expressiveness instead create a side channel usable for remote keystroke inference; even sanctioned, user-triggered gaze sampling yields sensitive attention data. Apple patched the specific keystroke-inference channel in visionOS~1.3 (CVE-2024-40865); the underlying outward-rendering design is unchanged. & \cite{99,110,60,100} \\
\addlinespace
Simulated social presence & EyeSight and Persona substitute a digitally rendered proxy for direct human contact; even where it improves perceived presence relative to no proxy, it remains short of face-to-face interaction, reframing isolation as a societal cost of mediated presence. & \cite{102} \\
\addlinespace
Hype-driven suppression of scrutiny & Promotional rhetoric surrounding AVP's launch and marketing shapes public and user discourse, constraining critical evaluation of the device's risks and overstating its everyday productivity value. & \cite{103,73,104} \\
\addlinespace
Price as structural exclusion & Purchase intention and adoption are sharply stratified by income, meaning the device, and the research literature built on it, systematically underrepresents lower- and middle-income users. & \cite{105} \\
\addlinespace
A conspicuously thin evidence base & Only 2 of 104 included papers were coded as centrally focused on misuse/risk/ethics, versus 68 on validated benefits, 7 on technical/ergonomic limitations, and 27 spanning more than one register; documented (as opposed to anticipated) harms, and topics such as workplace surveillance, remain largely unstudied. & This review (\Cref{tab:coding}) \\
\bottomrule
\end{tabular}
\end{strip}

\section{Discussion}\label{sec:discussion}

Read as a whole, the catalog in \Cref{sec:good,sec:bad,sec:ugly} says as much about the shape of AVP research to date as it does about AVP itself. Three patterns recur across all three sections and are worth naming explicitly.

First, the corpus comprises early adopters studying early adoption. 
The great majority of included papers are single-case reports, small pilot cohorts, or short controlled studies run by teams that already had access to a device that launched at \$3,499 and has since risen to \$3,699, a constraint several papers cite directly as limiting their generalizability. 

This is not a flaw specific to any one study; it is a structural feature of research on an expensive, single-vendor artifact in its first three years of availability, and it means the good cataloged in \Cref{sec:good}, however consistent across specialties, should be read as evidence of feasibility rather than of population-level effectiveness. 

Relatedly, our own coding rule (\Cref{sec:framework}) codes a study by its stated aim rather than its statistical outcome, which means a handful of null or mixed results (e.g. \cite{107}) are counted toward the good total above; this is a second, self-inflicted source of upward bias in the same direction as the early-adopter sampling problem, and future coding of this literature should consider outcome-weighted alternatives.

Second, the field has so far allocated its attention asymmetrically. 
Sixty-eight of 104 included papers were coded as centering on a validated benefit; two on ethical risk or misuse; a further seven on a technical/ergonomic limitation; and 27 as mixed, spanning more than one register (\Cref{tab:coding}). 
We do not read this as evidence that AVP is safer than other emerging technologies: published feasibility and usability studies are the wrong instrument for detecting structural harms in the first place, since those harms often look identical to ordinary technical immaturity until someone deliberately investigates whose interests a given design choice serves. 

What the asymmetry does tell us is where the marginal research contribution is largest: independent security research (of which GAZEploit is so far a rare example \cite{99}), bystander-centered study designs (rather than wearer-centered usability studies), and research that treats non-owners and non-users as a population worth studying, not just an absence in the sampling frame.

Third, several of the issues cataloged above do not cleanly resolve into bad or ugly using the kinds of evidence current studies collect. Price is a case in point: \Cref{sec:bad} treats it as an adoption barrier that constrains sample sizes across the corpus, one that later hardware generations may ease; \Cref{sec:ugly} treats the same figure as a stratifying mechanism, since purchase intention exceeds 75\% among high-income users while price sensitivity suppresses adoption in the middle-income segment, skewing who is represented in AVP research \cite{105}. 

Both readings are consistent with the same published data, because only Apple knows whether the current price reflects a declining cost curve or a deliberate premium-positioning choice. 

The price trajectory observed since launch is suggestive on this point: AVP held at \$3,499 through the October 2025 M5 hardware refresh, then rose to \$3,699 (256GB), \$3,899 (512GB), and \$4,199 (1TB) in June 2026, a change Apple attributed to rising memory and storage component costs rather than to a product refresh \cite{111}. 
A flat-then-rising trajectory is evidence against, not for, the ``declining cost curve'' reading that would ease the adoption barrier \Cref{sec:bad} describes; two years in, price has moved in the direction the structural-exclusion reading in this section anticipates, not the direction the technical-adoption-barrier reading in \Cref{sec:bad} would predict. 

Simulated social presence shows the same pattern: EyeSight and Persona measurably fall short of face-to-face interaction, which can be read as a limitation of current sensor and rendering technology or as a designed feature of mediating co-presence through a proprietary avatar system at all \cite{102}, and nothing in the usability data distinguishes a temporary shortfall from a permanent trade-off. 

The gaze-tracking privacy architecture raises a related tension: no continuous gaze-position API is exposed to third-party apps or researchers, yet the same architecture renders the wearer's eye movements into the outward-facing Persona stream, the exact channel that GAZEploit exploits to reconstruct a wearer's typed passwords \cite{99}. 

Apple's own response tells us how it handled this one instance, a targeted fix in visionOS 1.3 that suspends Persona during keyboard entry rather than a redesign of what gaze data Persona renders outward at all, but it does not tell us whether the underlying decision to render gaze onto Persona in every other context is a considered trade-off privileging avatar expressiveness or an oversight nobody has yet been paid to look for again; that remains an internal decision no published study can observe. 

What would resolve these cases is not another usability study but disclosure: a component-level roadmap from Apple, or an independent regulatory and security review, that converts an inference about intent into a fact.

These patterns suggest three concrete directions for future work: risk- and ethics-focused primary research, commissioned rather than incidental to studies designed for other purposes; bystander- and equity-centered study designs that do not require owning or wearing the device to participate; and continued attention, in future reviews of this literature, to findings like the ones above where the same published result is genuinely compatible with more than one reading, since resolving them depends on disclosure the studies themselves cannot supply.

\textbf{Limitations.} 

This review has its own methodological limits worth stating plainly. We searched a single database (Scopus); given how much of this corpus is ACM- and IEEE-published conference material, indexing gaps relative to the ACM Digital Library or IEEE Xplore directly are a real coverage risk rather than a formality, and a multi-database replication would be a natural next step. 
Screening was performed by two reviewers, one of whom screened every record and one of whom checked those decisions and ran a separate blind pass, but the eight borderline records were resolved by discussion, and although the blind pass would permit it, no inter-rater agreement statistic was computed.

The two papers retained as judgment calls under criterion~2 (Section~\ref{sec:incexc}), each comparing AVP with a no-headset baseline and one other headset, would be removed under a stricter single-device rule.

Screening was conducted at the title/abstract/document-type level rather than at the full-text level, consistent with the scope of an emerging-technology rapid review but a step short of a full systematic review. We did not conduct a quality appraisal of individual included studies (e.g., risk-of-bias scoring) or a meta-analysis, both of which the largely single-case and small-pilot nature of this literature would, in any case, make difficult to interpret. 

Finally, restricting the corpus to English-language records excludes any non-English primary research on AVP that may exist.

\section{Conclusion}\label{sec:conclusion}

We have reviewed the primary-focus academic literature on Apple Vision Pro published between its 2024 launch and 9 October 2026, screening 218 Scopus records down to 104 papers and organizing them using a good/bad/ugly structure adapted from Sergio Leone's tripartite framing in \textit{The Good, the Bad, and the Ugly}. 

The good is substantial and consistent: across surgery, diagnostic imaging, education, accessibility, design, and research instrumentation, AVP has repeatedly proven effective, often on a team's first attempt. 

The bad is equally consistent and, on the evidence, actively being engineered against: weight and comfort, input-method trade-offs, workload costs relative to established alternatives, and an immature software ecosystem recur across otherwise unrelated studies. 

The ugly is real but thin: gaze-based attack surfaces, income-stratified access, and hype rhetoric that outpaces critical scrutiny, and its thinness is itself a finding about where the field's attention has not yet gone.

Beyond the catalog, we have shown that some of the issues it documents do not resolve into a single verdict from the published evidence alone: price, simulated social presence, and the gaze-tracking privacy architecture each admit both a ``bad'' (technical-immaturity) and an ``ugly'' (structural-choice) reading that the data cannot adjudicate between. 
We treat this indeterminacy as a substantive finding rather than a shortcoming of our own analysis, and as a marker of where independent disclosure, not further usability testing, is what the field most needs next. 

Together, the catalog and this observation are offered not as a final verdict on whether Apple Vision Pro is good, bad, or ugly, but as an account of why that question, for this kind of technology, is one the published literature alone cannot yet settle, and a map of what would need to change, in the evidence and in the disclosure available to researchers, for it to.

\medskip\noindent\textbf{Data availability} The Scopus export underlying the screening reported here: \url{https://doi.org/10.5281/zenodo.23669705}, the screening decisions are available in section \cref{sec:method}.

\section*{Declaration of competing interest}
The authors declare no conflicts of interest.


\bibliographystyle{elsarticle-num}   
\bibliography{00Paper}

\end{document}